\documentclass[aps,reprint,superscriptaddress,amsmath,amssymb]{revtex4-1}

\usepackage{graphicx}
\usepackage{dcolumn}
\usepackage{bm}
\usepackage{hyperref}
\hypersetup{colorlinks=true, citecolor=blue, urlcolor=blue, linkcolor=blue}

\begin{document}

\title{Multi-channel Interference in Nonperturbative Multiphoton Pair Production by Gamma Rays Colliding}

\author{Zhaoyang Peng}
 \affiliation{Department of Physics, National University of Defense Technology, Changsha 410073, China}
 \affiliation{Department of Physics, Graduate School of China Academy of Engineering Physics, Beijing 100193, China}

\author{Huayu Hu}
 \email{huayu.hu.cardc@gmail.com}
 \affiliation{Hypervelocity Aerodynamics Institute, China Aerodynamics Research and Development Center, Mianyang 621000, Sichuan, China}
 \affiliation{Department of Physics, Graduate School of China Academy of Engineering Physics, Beijing 100193, China}

\author{Jianmin Yuan}
 \email{jmyuan@nudt.edu.cn}
 \affiliation{Department of Physics, National University of Defense Technology, Changsha 410073, China}
 \affiliation{Department of Physics, Graduate School of China Academy of Engineering Physics, Beijing 100193, China}

\date{\today}

\begin{abstract}
The electron-positron pair production in the collision of two intense gamma lasers is studied beyond the dipole approximation. The complete electron/positron momentum spectrum obtained by three-dimensional computation shows directly the multi-photon pair production channels and provides a sensitive test on the widely-used concept of the effective mass. Nonperturbative features are identified in the dependence of the total yield on the field intensity. In particular, various breaks in the momentum rings are observed and proved to be a manifestation of particle's energy gaps in the standing light wave induced by the Kapitza-Dirac scattering. A theoretical model of multi-level systems coupled by the Kapitza-Dirac scattering and pair production channels is established. Multi-channel interference effects and pair production suppression are revealed in these systems.

\end{abstract}

\maketitle

\section{Introduction}
The spontaneous electron-positron pair production (PP) is predicted to induce the vacuum breakdown in a constant electric field as the field strength exceeds the critical value $E_c=1.3\times10^{18}$~V/m \cite{PhysRev.128.2425,Sauter1931}, and also in heavy nuclei collision and extreme astrophysical events \cite{RUFFINI20101}. The relevant research develops quantum electrodynamics in the nonperturbative regime and provides significant insights to other branches of quantum field theory, such as the Adler-Bell-Jackiw anomaly \cite{PhysRevLett.117.061601} and string breaking in the strong interactions \cite{PhysRevD.20.179}. As experimentally motivated by the fast development of the strong-laser technology \cite{eli, xcels, Shen_2018}, abundance of study has been devoted to PP in spatially-homogeneous temporally-oscillating electric fields (OEF) \cite{RevModPhys.84.1177,PhysRevA.81.022122,PhysRevLett.102.150404,PhysRevLett.112.050402,PhysRevD.90.113004}, including interference effects to enhance PP \cite{PhysRevLett.104.250402, *PhysRevD.83.065028, PhysRevLett.108.030401}. However, generalization to spatially-inhomogeneous field poses a serious challenge in both analytical and numerical work \cite{RevModPhys.84.1177, Schneider2016, PhysRevLett.107.180403,PhysRevD.97.036026}.

In view of the fast-developing x-ray free-electron lasers and gamma ray \cite{RevModPhys.84.1177, PhysRevLett.116.185003, PhysRevLett.122.204802, An2018, chang2017brilliant}, we study PP in a standing wave composed of intense colliding gamma lasers. Note that the OEF can be regarded as a dipole approximation of the field around the electric anti-node of a standing wave. But this approximation is reasonable only if the laser wavelength is much larger than the characteristic length scale of the PP process, which requires $\omega\ll m$ and the nonlinear dimensionless parameter $\xi=|e|E/m\omega\gg 1$ \cite{PhysRevLett.102.080402}, where $E$ and $\omega$ are the electric field strength and frequency, $m$ and $e$ are the electron mass and charge, respectively, with natural units $\hbar=c=1$ used throughout. As we focus on the field regime $\xi \sim 1$ and $\omega \sim m$, the dipole approximation does not apply, and both the spatial inhomogeneity and the magnetic field are naturally involved.

Our study is based on the quantum electrodynamics with background fields \cite{unstable, GELIS20161, PhysRevLett.104.250402, *PhysRevD.83.065028, PhysRevLett.108.030401, PhysRevD.91.125026, PhysRevD.94.065024, PhysRevD.90.113004, PhysRevA.97.022515} and three-dimensional computation of Dirac equation. Previous one-dimensional calculations or calculations with specially chosen positron state for PP in such high-frequency standing waves have shown drastic differences compared to OEF in e.g., the momentum spectrum, the production yield, and the dependence of PP on laser polarization \cite{PhysRevA.97.022515,PhysRevD.96.076006,*PhysRevD.97.116001,PhysRevD.91.125026,PhysRevLett.102.080402}. In this paper both the multiphoton pair production channels of the Breit-Wheeler type ($n\gamma+n'\gamma' \rightarrow e^-e^+$) \cite{PhysRev.46.1087, reiss1962absorption, narozhnyi1964quantum, ritus1985quantum} and
the nonperturbative features of the process are shown. With the complete momentum spectrum, the widely used effective mass concept that there exists a mass-like quantity independent of the momentum in the strong field is put under a sensitive test. It is found that the Kapitza-Dirac scattering \cite{RevModPhys.79.929} plays an important role in the pair production, and in particular multi-channel destructive interference exists in this field which can suppress the pair production.

\section{Theoretical and Computational methods}
In quantum electrodynamics with background fields, the laser field is taken as a classical background field and the interaction between the quantized radiation field and the particles is neglected. This method is justified when the number of created pairs is small and the interaction between the particles as well as the feedback of the particles to the laser field can be neglected. The number of electrons produced in state $n$ is given by \cite{unstable,GELIS20161, PhysRevD.91.125026, PhysRevD.94.065024, PhysRevA.97.022515} (see Appendix \ref{appendixA} for details)
\begin{equation}\label{Num}
  \mathcal{N}_n=\left|\langle0,t_{in}|\hat{a}_n^\dag(t_{out})\hat{a}_n(t_{out})|0,t_{in}\rangle
\right|^2=\sum_{m}\left|G({}^+|{}_-)_{n;m}\right|^2,
\end{equation}
where $|0,t_{in}\rangle$ is the vacuum state at the instant $t_{in}$ when the field is not turned on yet and $\hat{a}_n^\dag(t_{out})$($\hat{a}_n(t_{out})$) is the creation (annihilation) operator of state $n$ at the instant $t_{out}$ when the field is turned off already. In the Dirac sea picture, $G({} ^+|{}_-)_{n;m}$ is equivalent to the external-field-induced transition amplitude from the negative-energy in-state $m$ to the positive-energy out-state $n$. Therefore, one has to solve the time-dependent Dirac equation for all negative-energy in-state as the initial states to obtain the momentum spectrum of particles.

We consider the collision of two laser beams propagating in the direction $\pm \bm{e}_z$ and polarized along $\bm{e}_y$. The laser beams have the equal intensity with $\xi_L=\xi_R=\xi/2$ (subscripts $L$ and $R$  denote the beams propagating along $\bm{e}_z$ and $-\bm{e}_z$, respectively) and the same frequency $\omega$. If the spatial dependence of the envelop is neglected, the laser field is approximated as a standing wave and the vector potential can be written as $\bm{A}=(m\xi/|e|)f(t)\sin(\omega t)\cos(\omega z)\bm{e}_y$, where the time profile $f(t)$ contains a plateau with duration $T$, and $sin^2$-like turn-on and turn-off phases of two laser cycles each. The Dirac equation in the field $\bm{A}$ is solved by the split-operator method \cite{PhysRevA.59.604, MOCKEN2008868, BAUKE20112454,FILLIONGOURDEAU20121403}, and the cost of computation is largely reduced by making use of the periodicity of $\bm{A}$ in the $z$ direction, seeing the details in Appendix \ref{appendixB}.

\section{Momentum spectrum and effective mass model}
\begin{figure}[t]
 \centering
 \includegraphics[scale=0.45,angle=0]{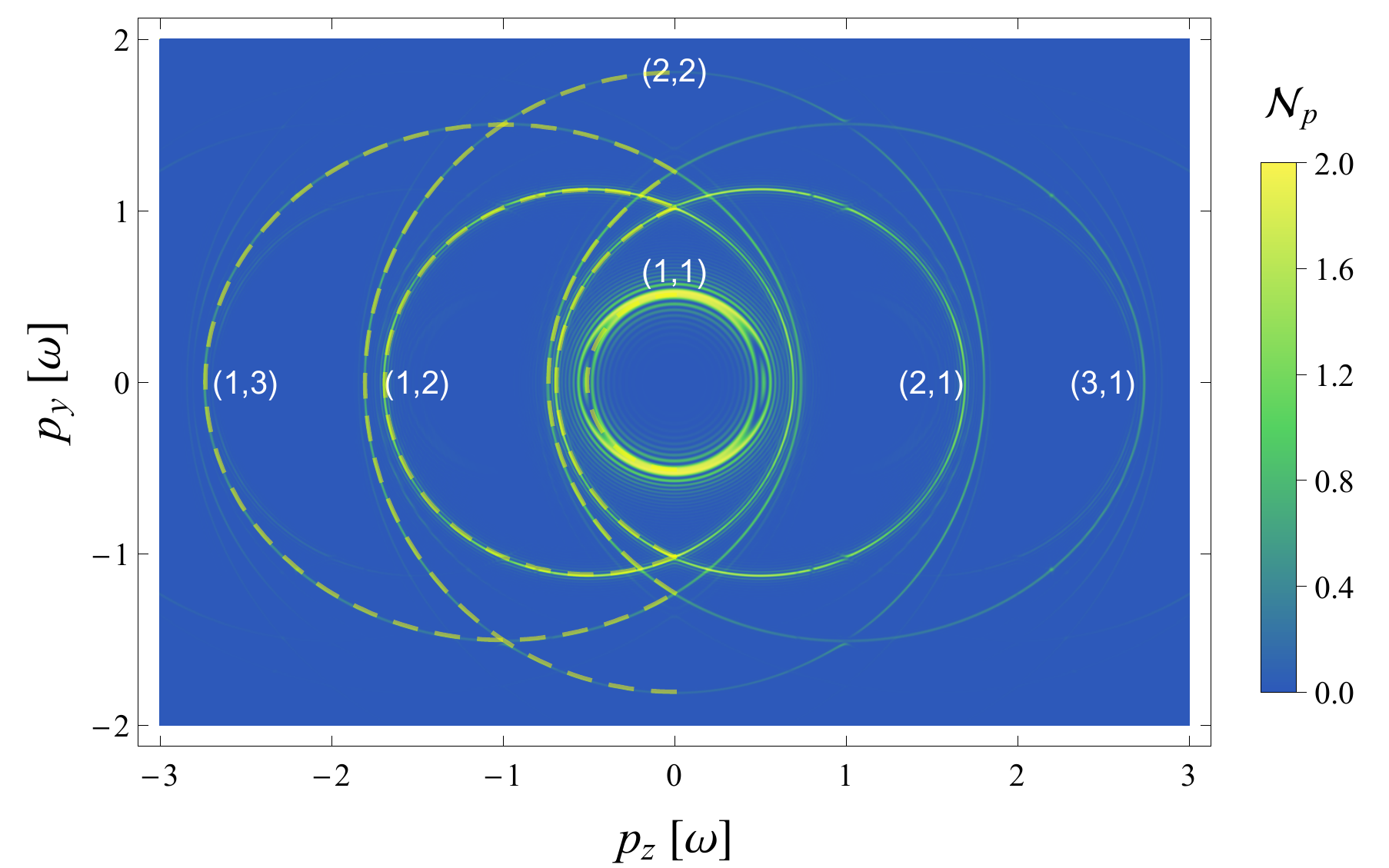}
 \caption{$(p_y, p_z)$ momentum spectrum of the electrons/positrons created with $p_x=0$ in the standing wave with $\omega=1.3$~m, $\xi=1$ and $T=40\tau$ with the laser cycle $\tau=2\pi/\omega$. Resonance rings are contributed from different $(n_L, n_R)$ channels as labeled. The analytical prediction (dashed line) for the position of each resonance ring obtained from Eqs.~\ref{resonance condition} uses an effective mass value extracted from the computed ring, as described in the text.}
 \label{ring}
\end{figure}

The two-dimensional momentum spectrum obtained in Fig.~\ref{ring} reveals characteristic resonance rings and clearly demonstrates the contribution of different multi-photon channels. There are circles centered at the origin contributed by the $(n_L, n_R)$ channels (abbreviation for $n_L\gamma_L+n_R\gamma_R\rightarrow e^-e^+$) with $n_L=n_R$ and ellipses centered at $(p_y=0,p_z=\omega(n_L-n_R)/2)$ contributed by channels with $n_L\neq n_R$. In comparison, there are only concentric resonance rings in the momentum distribution for the OEF case \cite{PhysRevA.81.022122,PhysRevD.90.113004}. The location of the resonance rings can be reproduced satisfactorily from the energy-momentum conservation relation with the dressed momentum assumption, as follows. Assume for an arbitrary positron(+)/electron(-) asymptotic state with momentum $\bm{p}^\pm$ outside the laser field, there exists a dressed momentum quantity $(\varepsilon_{q^\pm}, \bm{q}^{\pm})$ capable of describing the corresponding particle dynamics in the standing wave. Therefore, for the $(n_L, n_R)$ pair production channel, we may write the resonance condition according to the energy-momentum conservation as
\begin{equation}\label{resonance condition}
\begin{split}
   \varepsilon_{q^+}+ \varepsilon_{q^-}&=(n_L+n_R)\omega \\
   q_z^++q_z^-&=(n_L-n_R)\omega\\
   q_y^++q_y^-&=0.
\end{split}
\end{equation}
An effective mass can be defined as $m^*=\sqrt{\varepsilon_{q^\pm}^2-(\bm{p}^\pm)^2}$, and note that $m^*$ may be dependent on $\bm{p}$. We further assume that $\bm{q}^++\bm{q}^-\approx\bm{p}^++\bm{p}^-$. This approximation is reasonable when the ponderomotive energy $\sim m\xi^2\ll|\bm{p}^\pm|$, and also for the case $n_L\approx n_R$ with the symmetry considered. Therefore, reading a $(p_y,p_z)$ value from an arbitrarily chosen point on an arbitrary resonance ring in Fig. \ref{ring} and noting that the ring also represents the positron's momentum from the same channel, the Eq.~(\ref{resonance condition}) can be solved, the $n_L$ and $n_R$ value of this ring can be settled and the effective mass defined above can be determined for that point. In turn, with a single effective mass, $(p_y,p_z)$ relation can be drawn via Eq.~(\ref{resonance condition}) with different $n_L$ and $n_R$ values, shown as dashed lines in the figure. All the resonance rings are reproduced satisfactorily by a single $m^*\approx 1.12m$, indicating an almost complete independence of the so defined $m^*$ on $\bm{p}$. Hence, this demonstrates the predictive power and applicability of the effective mass model. And the positions of the momentum rings are given by
\begin{equation}\label{ring equation}
  \frac{4 n_L n_R(p_z-\omega \Delta n/2 )^2}{n^2(n_L n_R \omega^2-{m^*}^2)}+\frac{p_y^2}{n_L n_R \omega^2-{m^*}^2}=1,\\
\end{equation}
where $\Delta n=n_L-n_R$ and $n=n_L+n_R$.

\begin{figure}[t]
 \centering
 \includegraphics[scale=0.51,angle=0]{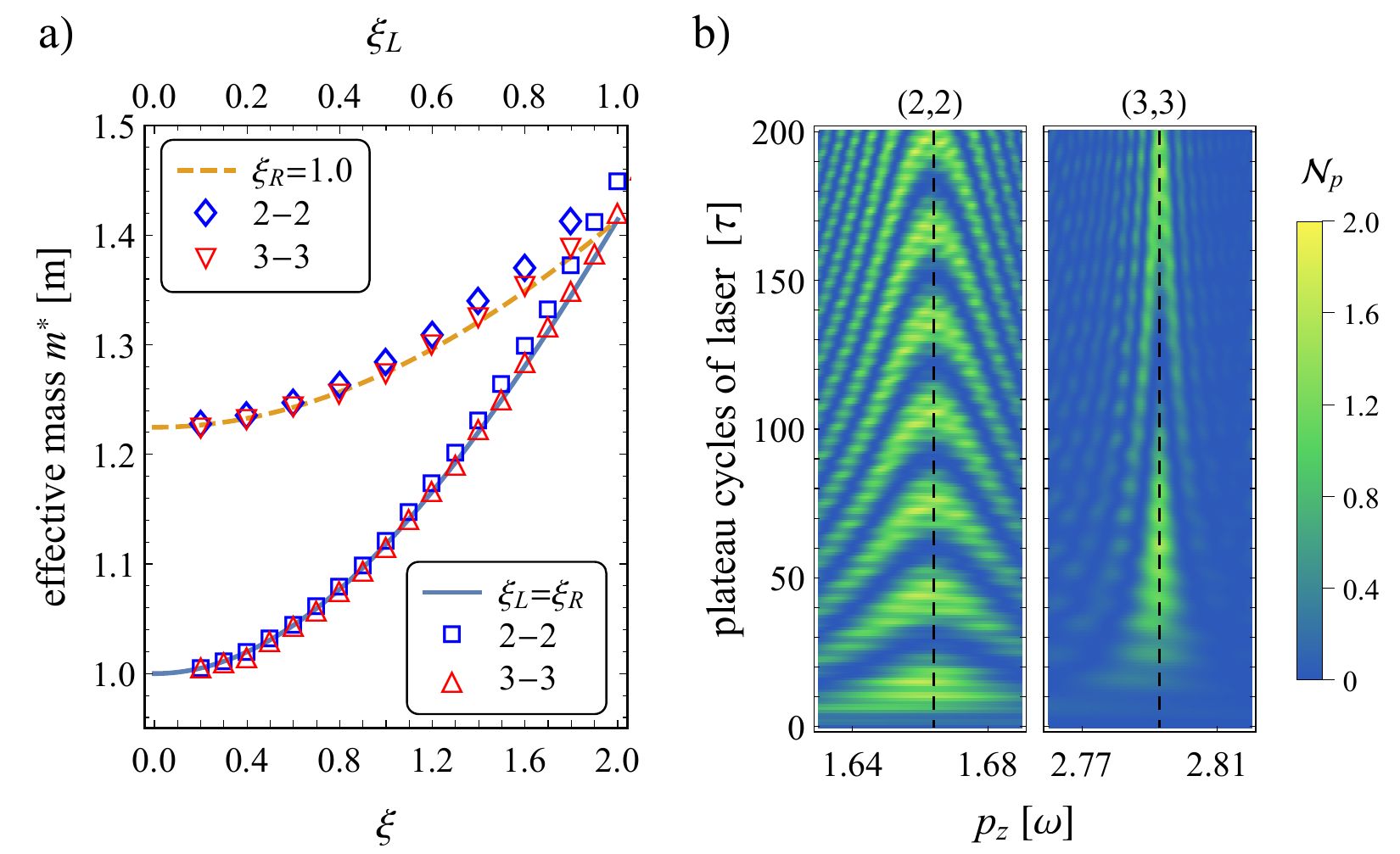}
 \caption{(a) Dependence of $m^*$ on $\xi$ for the standing-wave case ($\xi_L=\xi_R=\xi/2$) and on $\xi_L$ for fixed $\xi_R=1$ cases, with the same laser frequency $\omega=1.3$~m. The blue solid line and the yellow dashed line are results obtained using Eq.~(\ref{effeq}) for the respective cases. Blue squares, red triangles, blue diamonds and red inverted-triangles are $m^*$ values extracted from the $(2, 2)$ and $(3, 3)$ rings for the respective cases. (b) The resonance rings actually have a width that oscillates in time. The accurate position  of the point on the ring is determined among several Rabi oscillation cycles \cite{PhysRevLett.102.080402}. Shown here are the Rabi oscillations of the $(2, 2)$ and $(3, 3)$ channel at $p_x=p_y=0$ in a standing wave with $\omega=1.3m$ and $\xi=2$. The location of the resonance peak is $p_z=1.664\omega$ and $p_z=2.793\omega$ for the $(2, 2)$ and $(3, 3)$ channel, respectively, each marked by a vertical dashed line. These values are used to calculate the corresponding $m^*$ value at $\xi=2$ in (a).}
 \label{effmass}
\end{figure}

The dependence of  $m^*$ on the laser parameter (Fig.~\ref{effmass}) shows that the numerically extracted value of $m^*$ coincides well with the expression
\begin{equation}\label{effeq}
m^*=m\sqrt{1+\frac{\xi_L^2}{2}+\frac{\xi_R^2}{2}},
\end{equation}
which is derived analytically for a high energetic electron obliquely propagating in a standing wave \cite{PhysRevA.92.062105}. Remarkably, this coincidence is found to exist even in more general cases such as $\xi_L\neq \xi_R$.

\begin{figure}[t]
 \centering
 \includegraphics[scale=0.60,angle=0]{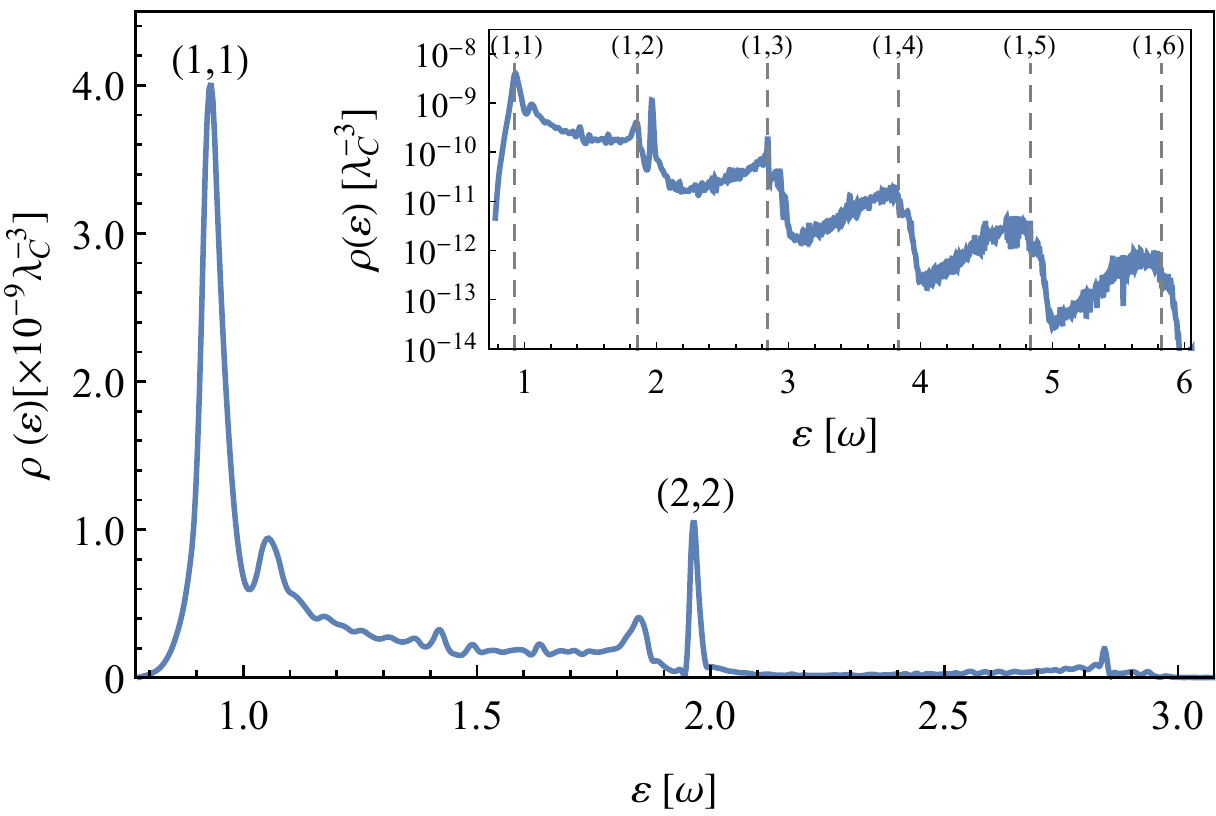}
 \caption{Energy spectrum of the electrons/positrons with $p_x=0$ produced from the same standing wave as in Fig. \ref{ring}. The two dominant peaks come from  $(1,1)$ and $(2,2)$ channel, respectively. The inset shows the step structure of the energy spectrum in a log scale. The vertical dashed lines marked as $(1,n)$ denote the maximum energies of the particle that can be created in the $1\gamma_{L/R}+n\gamma_{R/L}\rightarrow e^-e^+$ channels, respectively.}
 \label{espectrum}
\end{figure}

As can be directly obtained from the momentum spectrum Fig.~\ref{ring}, the energy $\varepsilon_{\bm{p}}=\sqrt{\bm{p}^2+m^2}$ of created electrons/positrons with $p_x=0$ is shown in Fig.~\ref{espectrum}. The dominant peaks come from $(n_L,n_R)$ production channels with $n_L=n_R$ because the electrons/positrons created by such kind of a channel have a single energy value. For example, according to Eq.~(\ref{resonance condition}) the dressed energies of the electrons/positrons produced in the laser field by the $(1,1)$ and $(2,2)$ channels should be $1\omega$ and $2\omega$ respectively. Due to the effective mass effect discussed above, the energy of the particle detected in the vacuum should be lower than its dressed energy in the laser field, and thus the locations of the two peaks shift to lower values correspondingly. Although the dominant peaks have also been identified in previous work \cite{PhysRevA.97.022515}, the step structure of the plateaus shown in the inset has not been discussed before. They are actually contributed by $(n_L,n_R)$ channels with $n_L\neq n_R$, since the energy of the electrons/positrons created by the $(n_L,n_R)$ channel distributes between the $(n_L,n_L)$ and $(n_R,n_R)$ peaks. The cut-off of each plateau coincides well with the maximum particle energy that can be produced by the channels $1\gamma_{L/R}+n\gamma_{R/L}\rightarrow e^-e^+$ with $n=1,2,\cdots$.

\section{Breaks in the momentum ring and multi-channel interference}
\begin{figure}[h]
 \centering
 \includegraphics[scale=0.5,angle=0]{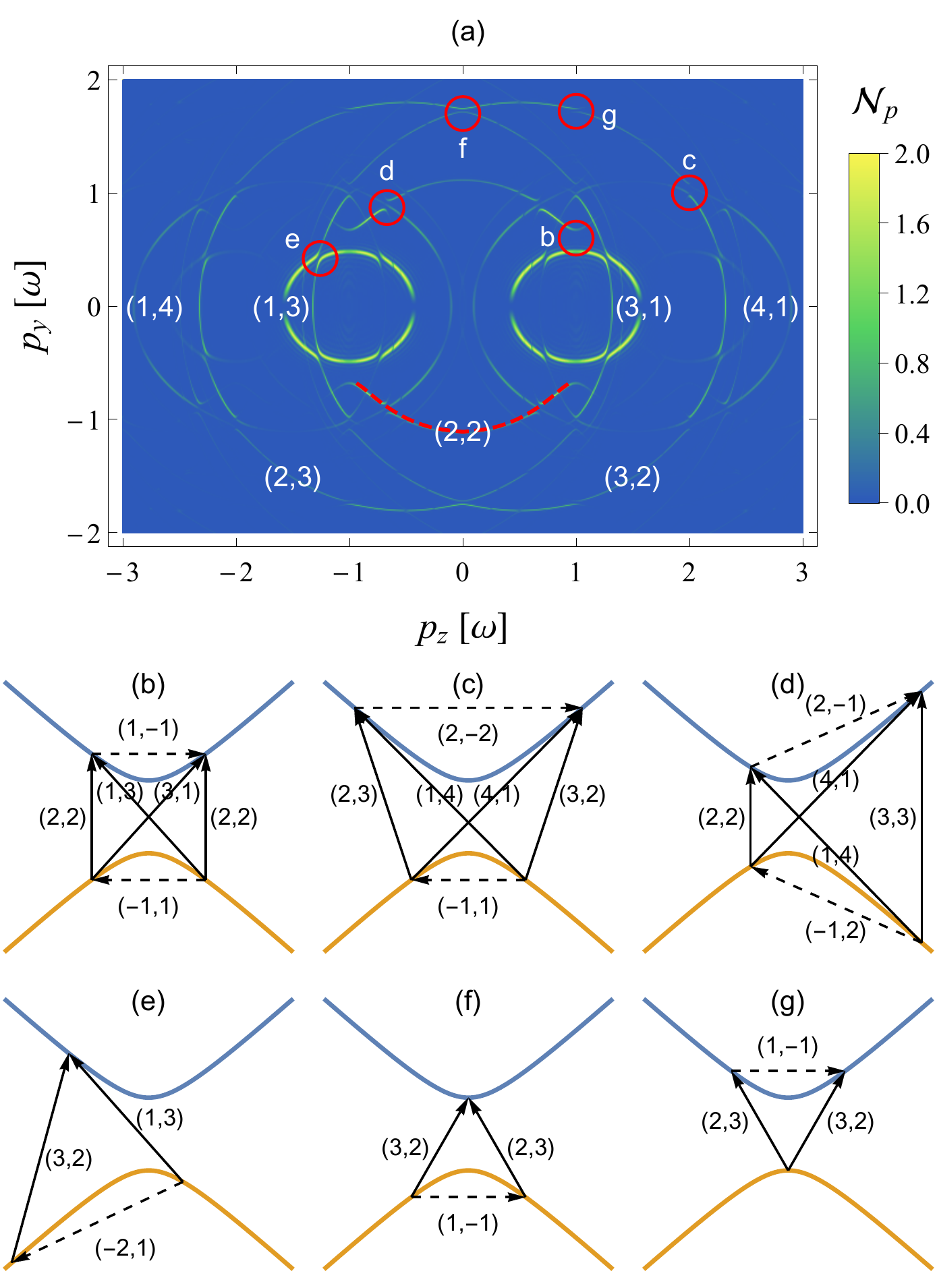}
 \caption{(a) $(p_y, p_z)$ momentum spectrum for the electrons/positrons created with $p_x=0$ in the standing wave of $\omega=0.67m$, $\xi=1$ and $T=60\tau$. (b)$-$(g) The states related to the breaks marked as b$-$g in (a) are illustrated on the energy-momentum curve. The pair production channels (solid arrow) and KD scattering channels (dashed arrow) among the states are shown.}
 \label{coupling}
\end{figure}

Breaks are found in the momentum resonance rings. Some take place at the crossings of two rings as the marked b$-$f cases in Fig.~\ref{coupling}. The positive-energy state at the crossing can be populated by the corresponding two different pair production channels $(n_1, m_1)$ and $(n_2, m_2)$, and according to the energy-momentum conservation the corresponding initial negative-energy state of the $(n_1, m_1)$ channel can transit to that of the $(n_2, m_2)$ channel by the $(n_1-n_2, m_1-m_2)$ Kapitza-Dirac (KD) scattering of absorbing ($+$ sign)/emitting ($-$ sign) $|n_1-n_2|$ ($|m_1-m_2|$) photons from the left (right) beam \cite{PhysRevLett.109.043601,*PhysRevA.88.012115}, as shown in Fig.~\ref{coupling}(b)$-$(f). Besides, KD scattering can also be present between the positive-energy states, see Fig.~\ref{coupling}(b)$-$(d)$\&$(g).  Note that the break g is not at any crossing. It can be seen that the states coupled by multiphoton pair production channels and KD scattering channels form the three- or four-level systems.

If the coupling via KD scattering is much stronger than via the pair production, such as in our cases (b$-$g) where the KD process requires fewer photons and thus has a larger scattering matrix amplitude for the laser parameter $\xi\sim 1$, the KD scattering results in energy shifts of the states, and thus the breaks and distortions of the resonance rings. Take the break b on the $(2,2)$ ring as an example. The two positive-/negative-energy states with $\bm{p}_\pm=(p_x,p_y, \pm \omega+\Delta p)$ can be seen respectively as a two-level system coupled by the KD scattering $(1,-1)$, which results in the energy shifts $\varepsilon'_{\pm}=(\varepsilon_++\varepsilon_-)/2 \pm \sqrt{(\varepsilon_+ - \varepsilon_-)^2+\Omega^2}/2$ \cite{QM}, where $\varepsilon_{\pm}=\sqrt{\bm{p}^2_\pm+{m^*}^2}$ and $\Omega$ is the Rabi frequency between the two states. With $\Delta p\ll \omega$, $\Omega$ can be approximated as the Rabi frequency $\Omega_{(1,-1)}$ between the two degenerate states $(p_x,p_y, \pm \omega)$ which could be numerically computed. Combining this energy shift and the resonance condition (\ref{resonance condition}), the break and distortions of the $(2, 2)$ ring can be reproduced satisfactorily, shown as the dashed line in Fig.~\ref{coupling}(a).  The energy shift could also be seen as the energy gap of the particles in the spatially periodic standing wave like in a crystal. But the energy gap could only be analytically obtained under very special boundary conditions, such as $p_y=p_x=0$ \cite{becker1979motion,*becker1979motion2}.  Here we prove that the energy gap could be obtained by numerically calculating the Rabi frequency between the corresponding states of the KD scattering.

\begin{figure}[h]
 \centering
 \includegraphics[scale=0.17,angle=0]{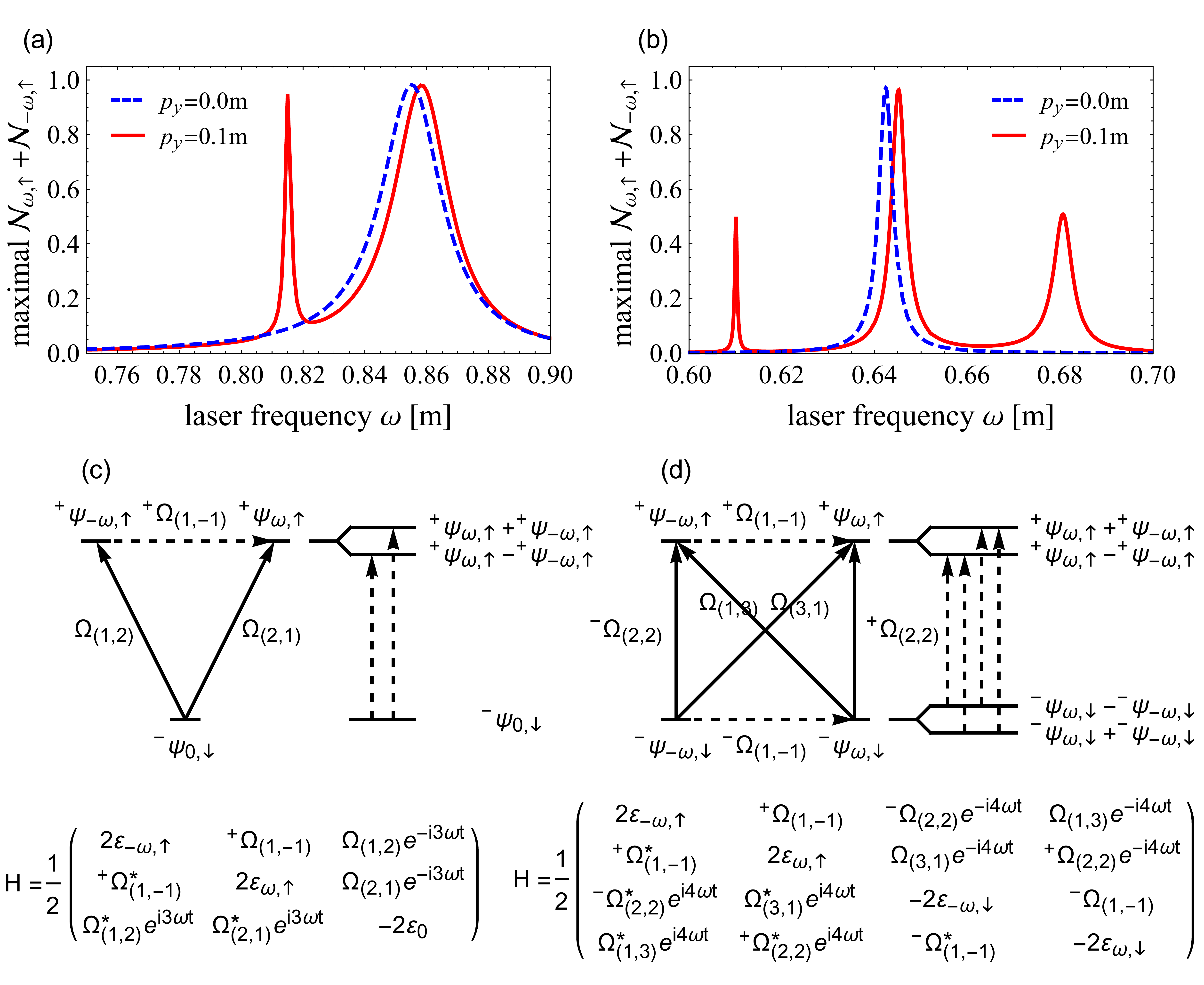}
 \caption{The maximal occupation value of the positive-energy states with $p_z=\pm \omega$ during 600 laser cycles with $\xi=1$ from a negative-energy states with (a) $p_z=0$ and (b) $p_z=-\omega$.
 The corresponding multi-level system and the effective Hamiltonian of each system are shown in (c) and (d). See details in the text.
 }
 \label{coherence}
\end{figure}

To illustrate the coherence of the multi-channel dynamics, the three/four-level systems like those in Fig.~\ref{coupling}(g) and (b) are separately studied as shown in Fig.~\ref{coherence}(a,c) and (b,d), where the maximal occupation value of the upper states or equivalently the maximum expectation number of the created electron during 600 laser cycles are manifested for different laser frequencies with the initial state set to be the lower state of each system, that is the negative-energy state $|^-\psi_{0,\downarrow}\rangle$ with $p_z=0$ for the three-level system and $|^-\psi_{-\omega,\downarrow}\rangle$ with $p_z=-\omega$ for the four-level system ($\uparrow/\downarrow$ denotes the spin along the $\pm x$). Besides, for each system two cases are investigated with $p_y=0$ and $p_y=0.1m$, respectively, while keeping $p_x=0$. Notice that the negative- and positive-energy state could be coupled only when they have opposite spin in the case of $p_x=0$. For the $p_y=0.1m$ cases, two resonance peaks are found for the three-level system \cite{PhysRevLett.102.080402}, while three peaks are found for the four-level system. This can be explained by the energy splitting of the degenerate upper states (see Fig.~\ref{coherence}(c)) as well as that of the degenerate lower states (see Fig.~\ref{coherence}(d)) induced by the KD scattering.

What is particularly interesting is the missing of some resonance peaks in the $p_y=0$ cases \cite{PhysRevD.91.125026}. To analyze this, an effective Hamiltonian matrix similar to that in quantum optics can be set up, shown in Fig.~\ref{coherence}, with the effective mass being used throughout which reflects the strong field effect. The diagonal elements of the matrix are the bare state energies, and the off-diagonal elements are the coupling amplitudes among the states as marked in Fig.~\ref{coherence}(c,d). It can be derived that the coupling amplitude is proportional to the QED scattering matrix term of the corresponding process, for example the Rabi frequency $^+\Omega_{(1,-1)}$ is proportional to the KD scattering matrix term $\langle^+\psi_{\omega,\uparrow}, \gamma_R|\hat{S}|^+\psi_{-\omega,\uparrow}, \gamma_L\rangle$ with $\gamma_{L/R}$ being the photon from the left/right beam, $\Omega_{(2,1)}$ is proportional to the pair production matrix term $\langle p^-_z=\omega, p^+_z=0 |\hat{S}|2\gamma_L, \gamma_R\rangle$, and so on.

For the three level system in Fig.~\ref{coherence}(c) in the case $p_y=0$, there is $\Omega_{(2,1)}=\Omega_{(1,2)}$ due to the invariance of the S-matrix under parity transformation, the odd number of participating photons, and the fact that the electron and positron are different parity eigenstates. Thus the state $|^+ \psi_{-\omega, \uparrow}\rangle- |{}^+\psi_{\omega,\uparrow}\rangle$ can not be stimulated, which results in the missing of the peak around $\omega=0.82m$ in Fig.~\ref{coherence}(a). This state resembles a dark state of the system in the laser field.

For the four level system in Fig.~\ref{coherence}(d) in the case $p_y=0$, as similarly can be proven by parity transformation, there are $^+\Omega_{(2,2)}=-^-\Omega_{(2,2)}$ and $\Omega_{(3,1)}=-\Omega_{(1,3)}$. Therefore, the transition from $|^- \psi_{-\omega,\downarrow}\rangle\pm |{}^- \psi_{\omega,\downarrow}\rangle$ to $|^+\psi_{-\omega,\uparrow}\rangle\pm |{}^+\psi_{\omega,\uparrow}\rangle$ are inhibited due to the destructive interference. Besides, due to the invariance of the S-matrix under charge conjugation and the existence of a minus sign between the negative-energy electron propagator and the positron propagator, there is $^+\Omega_{(1,-1)}=-^-\Omega_{(1,-1)}$, and the transition $|^- \psi_{-\omega,\downarrow}\rangle\pm |{}^-
\psi_{\omega,\downarrow}\rangle$ to $|^+ \psi_{-\omega,\uparrow}\rangle\mp |{}^+ \psi_{\omega, \uparrow}\rangle$ results in the remaining middle resonance peak. Previously only the multiple slit interference in the time domains studied for pair production in a homogenous time-dependent electric field \cite{PhysRevLett.104.250402, PhysRevD.83.065028, PhysRevLett.108.030401}, here we illustrate a different interference mechanism among different pair production channels in a space- and time-dependent field.

\section{Total yield of the pairs}

\begin{figure}[h]
 \centering
 \includegraphics[scale=0.6,angle=0]{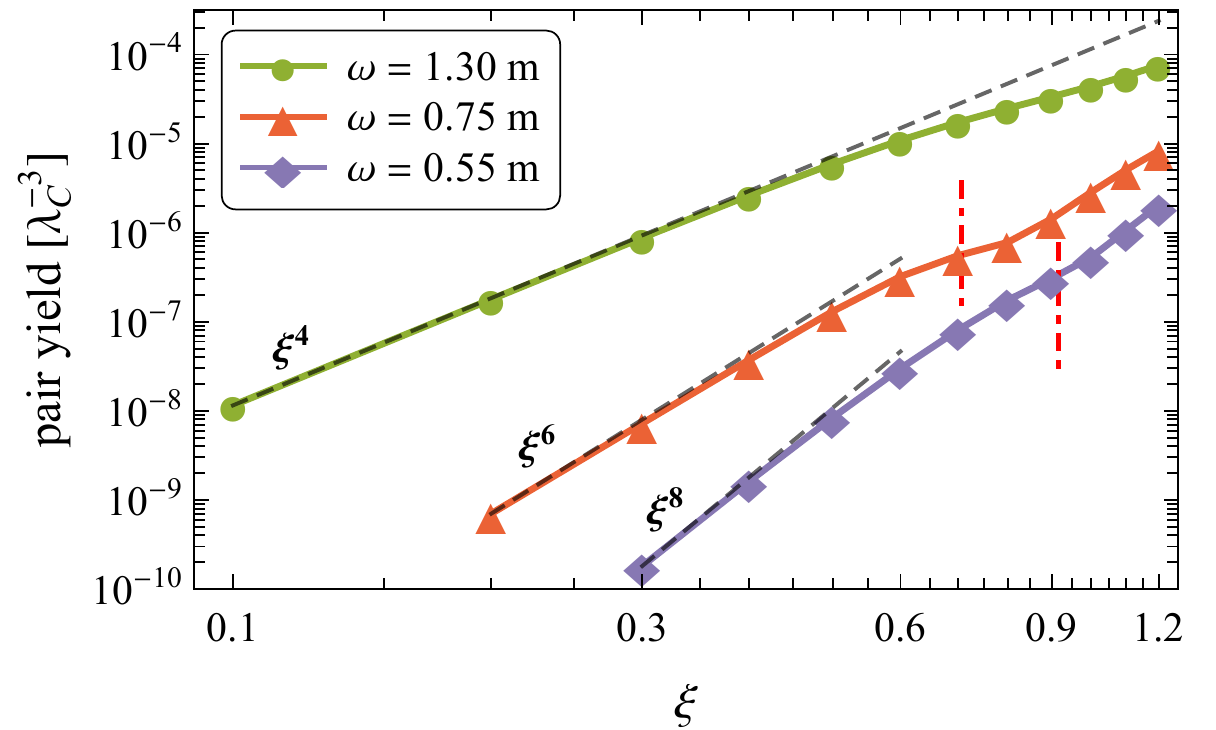}
 \caption{Dependence of total production yield per volume ($ \lambda_C=2\pi/m$ is the Compton wavelength) on $\xi$ in a standing wave with $T=10\tau$ and $\omega=1.30m$, $0.75m$ and $0.55m$. The grey dashed lines denote the function $\xi^{2n}$ where $n$ is the minimum number of photons absorbed for the pair production to take place obtained by the threshold condition. The vertical red dash-dotted lines marks the threshold where channel closing occurs.}
 \label{yield}
\end{figure}

Figure \ref{yield} shows the dependence of the total production yield per volume on $\xi$ for various laser frequencies. Equation (\ref{resonance condition}) results in the threshold condition $n_L\times n_R \geq (m^*/\omega)^2$ for the $(n_L, n_R)$ channel. As $\xi\ll 1$, the perturbative feature of the process is manifested, that for a $n$-photon process ($n=n_L+n_R$), the yield increases as $\xi^{2n}$. This also provides a verification of the numerical computation. As $m^*$ increases with $\xi$, the 3-photon pair production channel closes at $\xi=0.71$ for $\omega=0.75m$ and the 4-photon channel closes at $\xi=0.92$ for $\omega=0.55m$, showing an inflection in the lines. With $\omega=1.30m$, the deflection of the yield curve from the $\xi^4$ dependence manifests the onset of the non-perturbative feature when $\xi\rightarrow 1$, similar to that found for nonlinear-Compton scattering \cite{PhysRevLett.114.195003} and pair production in relatively low-frequency laser fields \cite{PhysRevLett.105.080401}.

\section{Summary}$e^-e^+$ pair production in an intense standing wave is studied. Both the Breit-Wheeler and the nonperturbative feature are identified. The concept of the effective mass is sensitively justified and an analytical model is obtained.  The KD scattering of particles could induce gaps in the particle's energy level which can lead to the breaks and distortions of the resonance rings in the electron's momentum spectrum. Accordingly, a strong field version of the multi-level system model like that in quantum optics is established, where the systems are composed of the states coupled by KD scattering and the multiphoton pair production channels. The coherent dynamics among these channels can lead to dark state and destructive interference which result in the suppression of the pair production. The coexistence of the strong field effect and the multi-channel interference can in principle be measured by future strong field pair production experiments with high resolution of the particle states.

\begin{acknowledgments}
H.H. acknowledges the support by the National Natural Science Foundation of China under Grant No. 11774415. J.Y. acknowledges the support from Science Challenge Project No. TZ2018005 and the National Natural Science Foundation of China under Grant No. 11774322.
\end{acknowledgments}

\appendix

\section{\label{appendixA}Theory of pair production in an intense classical field}

The real-time simulation of pair production is carried out by treating the strong coherent laser field as a classical field and using the mode-function method to obtain the dynamics of the fermion fields. In quantum electrodynamics with background fields, the total electromagnetic field is then treated as a combination of this intense classical field and a quantized radiation field of other photon modes \cite{unstable}. For pair production by an intense electromagnetic field from the vacuum, the interaction between the quantized radiation field and the particle as well as the feedback of the particles to the external field can be neglected, when the number of created pairs is small. However, the dynamics of the fermion field are difficult to address. We resort to the mode-function method by which one has to solve the Dirac equation in the external field for all mode functions to obtain the evolution of the fermion field operators \cite{GELIS20161}. Then the rate of particle production can be calculated by the Bogoliubov transformation between the IN and OUT creation and annihilation operators of the fermion particles \cite{GELIS20161, PhysRevLett.104.250402}. The corresponding numerical computation can be very time consuming and thus is usually applied for 1+1 dimensional problems. Our simulation is conducted in 3+1 dimensions by using the symmetry of the laser field to reduce the computation cost considerably, as illustrated below.

In the Heisenberg picture, the spinor field operator $\hat{\psi}(\bm{r},t)$ obeys the Dirac equation
\begin{equation}\label{Dirac_equation}
    i\frac{\partial}{\partial t}\hat{\psi}(\bm{r},t)= \hat{H}_D \hat{\psi}(\bm{r},t) =[\bm{\alpha} \cdot \left(-i\bm{\nabla}-e\bm{A}(\bm{r},t)\right)+m\beta] \hat{\psi}(\bm{r},t),
\end{equation}
where $\bm{\alpha}$ and $\beta$ are the Dirac matrices, and $\bm{A}(\bm{r},t)$ is the vector potential of the classical external field with the temporal gauge condition $A_{0}=0$ and the natural units $\hbar=c=1$ used throughout.  If the external field vanishes beyond a finite time interval $(t_{in},t_{out})$, the particle field $\hat{\psi}(\bm{r},t)$ at $t_{in}$ and $t_{out}$ moments can be expanded as
\begin{subequations}
  \begin{align}
    \hat{\psi}(\bm{r},t_{in})&=\sum_n {}_+\varphi_n(\bm{r})\hat{a}_n(t_{in})+{}_-\varphi_n(\bm{r})\hat{b}_n^\dag(t_{in}) , \label{}\\
    \hat{\psi}(\bm{r},t_{out})&=\sum_n {}^+\varphi_n(\bm{r})\hat{a}_n(t_{out})+{}^-\varphi_n(\bm{r})\hat{b}_n^\dag(t_{out}) ,\label{}
  \end{align}
\end{subequations}
where the mode functions ${}_\pm \varphi_n$(${}^\pm \varphi_n$) are the complete and orthogonal eigenstates of $\hat{H}_D$ at the instant $t_{in}$ ($t_{out}$), with $\pm$ respectively denoting the positive- and negative-energy solution, and the quantum number $n$ includes the momentum and the spin $n =\{\bm{p}, s\}$. $\hat{a}^\dag$, $\hat{a}$ and $\hat{b}^\dag$, $\hat{b}$ denote the creation and annihilation operators for the free electron and positron respectively, which satisfy the equal-time anti-commutation relation:
\begin{equation}\label{}
  \{\hat{a}_{n},\hat{a}^\dag_{m}\}=\{\hat{b}_{n},\hat{b}^\dag_{m}\}=\delta_{nm}\\
\end{equation}
 with all other equal-time anticommutators vanishing. The initial and the final vacuum states, that $|0,t_{in}\rangle$ and $|0,t_{out}\rangle$, are defined by $\hat{a}_n(t_{in}) |0,t_{in}\rangle=\hat{b}_n(t_{in}) |0,t_{in}\rangle=0$ and $\hat{a}_n(t_{out}) |0,t_{out}\rangle=\hat{b}_n(t_{out}) |0,t_{out}\rangle=0$, respectively. By solving Eq.~(\ref{Dirac_equation}), the electron's/positron's creation and annihilation operators at different instants $t_{in}$ and $t_{out}$ are connected by the linear canonical transformation (Bogoliubov transformation) \cite{unstable, GELIS20161, PhysRevD.91.125026, PhysRevD.94.065024}:
\begin{subequations}
\begin{align}
  \hat{a}_n(t_{out})&=\sum_m G({}^+|{}_+)_{n;m} \hat{a}_m(t_{in})+G({}^+|{}_-)_{n;m}\hat{b}_m^\dag(t_{in}) ,\label{a_operator}\\
  \hat{b}_n^\dag(t_{out})&=\sum_m G({}^-|{}_+)_{n;m} ,
  \hat{a}_m(t_{in})+G({}^-|{}_-)_{n;m}\hat{b}_m^\dag(t_{in}) ,
\end{align}
\end{subequations}
with $G({}^\pm|{}_\pm)_{n;m}$ defined as
\begin{equation}\label{green_matrices}
\begin{split}
  G({}^\pm|{}_\pm)_{n;m}&=\langle ^\pm\varphi_n |
  \hat{U}(t_{out},t_{in})
  |{}_\pm\varphi_m \rangle ,\\
  \hat{U}(t',t)&=\hat{T} \exp\left[ -i \int_{t}^{t'} \hat{H}_D (\tau)d\tau \right] ,\\
\end{split}
\end{equation}
where $\hat{U}$ is the unitary evolution operator and $\hat{T}$ is the time order operator. In the Dirac sea picture, $G({} ^+|{}_-)_{n;m}$ is equivalent to the external-field-induced transition amplitude from the negative-energy in-state $m$ to the positive-energy out-state $n$.

Therefore the average number of electrons produced in the state $n$ is given by

\begin{equation}\label{occupation_number}
  \mathcal{N}_n
  =\langle0,t_{in}|\hat{a}^\dag_n(t_{out})\hat{a}_n(t_{out})|0,t_{in}\rangle
  =\sum_m\left|G({}^+|{}_-)_{n;m}\right|^2, \\
\end{equation}
and the total yield of created electrons is
\begin{equation}\label{total_yield}
  \mathcal{N}=\sum_{m;n}\left|G({}^+|{}_-)_{n;m}\right|^2.
\end{equation}

Therefore, one should start at $t_{in}$ with an arbitrary negative-energy plane wave with momentum $\bm{p}$ and spin $s$ which can be interpreted as a positive-energy antiparticle with momentum $-\bm{p}$ and spin $-s$, then make it evolve in the external electromagnetic field, and finally project it on positive-energy plane waves at $t_{out}$. This procedure should be repeated for all negative-energy in-states as the initial state to obtain the momentum spectrum of particles. Since a discretized box containing the number of $N^3$ lattice sites supports a discretized momentum space of $N^3$ momentum modes, the computational cost of the three-dimensional problem is proportional to $N^6N_t$, where $N_t$ is the number of time steps.

\section{\label{appendixB} Numerical method for simulating pair production in an intense standing-wave field}
According to Eq.~(\ref{green_matrices}, \ref{occupation_number}, \ref{total_yield}), to compute the total yield one should solve the time-dependent Dirac equation for all initial states $_{-}\varphi_n$. Generally speaking, the computation is very time consuming for a temporal and spatial dependent field. However, the cost can be reduced when the symmetry of the external field is made use of. The standing wave field considered here is formed by two linearly polarized (polarization in the $\bm{e}_y$ direction) laser beams  propagating in opposite directions $\pm\bm{e}_z$ with the equal intensity $\xi/2$ and the same frequency $\omega$. The vector potential reads
\begin{equation}\label{}
  \bm{A}=\frac{m\xi}{|e|}f(t)\sin(\omega t)\cos(\omega z)\bm{e}_y,\\
\end{equation}
where the time profile $f(t)$ is
\begin{equation}\label{}
   f(t)=
    \begin{cases}
     \sin^2\frac{\pi(t-2\tau)}{4 \tau} &\text{if $-2\tau\leq t \leq 0$} \,,\\
     1 &\text{if $0\leq t \leq T$}\,, \\
     \cos^2\frac{\pi(t-T)}{4 \tau} &\text{if $T \leq t \leq T+2\tau $}\,,\\
     0 &\text{otherwise}\,,\\
    \end{cases}
\end{equation}
which contains a plateau $T$ and two cycles $2\tau$ for turn-on and turn-off with $\tau=2\pi/\omega$. Notice that the continuous translational invariance of the vector potential along $\bm{e}_x$ and $\bm{e}_y$ leads to conserved canonical momentum $p_x$ and $p_y$. And the periodicity of the vector potential along $\bm{e}_z$ results in a transition selection rule that a momentum eigenstate $_-\varphi_{\bm{p},s}$ could only jump to momentum eigenstates $^{\pm}\varphi_{\bm{p}+ l\bm{k},s'}$ with integer $l=0,\pm1,\pm2, \cdots$ and $\bm{k}=\omega\bm{e}_z$ \cite{PhysRevD.91.125026}. Therefore, Eq.~(\ref{a_operator}) can be reduced to
\begin{widetext}
\begin{equation}\label{}
  \hat{a}_{\bm{p},s}(t_{out})=\sum_{s'} \sum_l G({}^+|{}_+)_{\bm{p},s;\bm{p}+l\bm{k},s'} \hat{a}_{\bm{p}+l\bm{k},s'}(t_{in})+G({}^+|{}_-)_{\bm{p},s;\bm{p}+l\bm{k},s'} \hat{b}^\dag_{\bm{p}+l\bm{k},s'}(t_{in}) \,.
\end{equation}
\end{widetext}
The average number of produced electrons with the momentum $\bm{p}$ is
\begin{equation}\label{}
 \mathcal{N}_{\bm{p}}=\sum_{l,s,s'} \left|G({}^+|{}_-)_{\bm{p},s;\bm{p}+l\bm{k},s'}\right|^2.\\
\end{equation}
The Green matrices $G({}^+|{}_-)_{\bm{p},s;\bm{p}+l\bm{k},s'}$ could be obtained by evolving the initial states ${}_{-}\varphi_{\bm{p}+l\bm{k},s}$ within a subspace spanned by free spinors with momentum $\bm{p}+l \bm{k}$ \cite{PhysRevD.91.125026}. If one wants to get the average number of electrons/positrons of the definite momentum $\bm{p}$, one could evolve the negative-energy plane wave ${}_{-}\varphi_{\bm{p}+l\bm{k},s}$  of different $l$ and $s$ within such a subspace. Therefore, the complete momentum spectrum could be obtained by calculating the  average number of electrons/positrons of all momenta.

However, the complete momentum spectrum could be obtained more efficiently by constructing initial wave packets $\phi_{r,s}(t_{in})$  as
\begin{equation}\label{initial}
  \phi_{r,s}(t_{in})=\int_{r\omega}^{(r+1) \omega}dp'_{z}
  \int_{-r_m\omega}^{r_m\omega} dp'_{x}
  \int_{-r_m\omega}^{r_m\omega} dp'_{y}
  \left({}_{-}\varphi_{\bm{p}',s}(\bm{r})\right)\,, \\
\end{equation}
where the integral range is $\left[ r\omega, (r+1)\omega \right)$ with integer $r=-r_m, ..., r_m-1$ for $p'_z$  and $[-r_m\omega,r_m\omega)$ for both $p'_x$ and $p'_y$, where $r_m\omega$ is the truncation of the momentum space which is imposed by the spatial grid step $\pi/(r_m\omega)$ and should ensure the convergence. For example, taking $\omega=1$MeV and the lattice step in the $z$ direction $\Delta z=0.1\pi/$MeV, there is $r_{m}=\pi/(\Delta z \omega)=10$. And the integral turns into the summation in the discretized momentum space. Notice that
\begin{widetext}
\begin{equation}\label{}
  \sum_{r,s,s'}\left|
  \langle {}^+\varphi_{\bm{p},s} |
  \hat{U}(t_{out},t_{in})
  | \phi_{r,s'}(t_{in}) \rangle \right|^2=\sum_{l,s,s'}\left|G({}^+|{}_-)_{\bm{p},s;\bm{p}+l\bm{k},s'}\right|^2 =\mathcal{N}_{\bm{p}}, \\
\end{equation}
\end{widetext}
where the term on the left of the first equal sign indicates that one should evolve initial wave packets $\phi_{r,s}(t_{in})$ with different $r$ and $s$ and then project them on the positive-energy plane wave $^{+}\varphi_{\bm{p},s}$ at the final time. This equality is valid because each initial wave packet $\phi_{r,s}(t_{in})$ contains only one momentum component which could jump to the state $^{+}\varphi_{\bm{p},s}$ and there are no interference terms, according to Eq.~(\ref{initial}) and the transition rule illustrated above. Therefore, one just need to evolve these wave packets $\phi_{r,s}(t_{in})$ instead of evolving the negative-energy plane waves $_{-}\varphi_{\bm{p},s}$ one by one. In this way the computation cost has been reduced from $N^6N_t$ to $ 2r_{m} N^3N_t$.

The split-operator method \cite{PhysRevA.59.604, MOCKEN2008868, BAUKE20112454, FILLIONGOURDEAU20121403} is used to solve the Dirac equation to obtain the evolution of the initial wave packets. The unitary evolution operator in Eq.~(\ref{green_matrices}) satisfies
the composition condition
\begin{equation}\label{unitary_decom}
  \hat{U}(t_{out},t_{in})= \prod^{N_t}_{n=1} \hat{U}(t_{in}+n \Delta t,t_{in}+(n-1)\Delta t),\\
\end{equation}
where time steps $\Delta t =(t_{out}- t_{in}) / N_t$. At each step the Hamiltonian $\hat{H}_D$ is decomposed as a sum of two operators $\hat{K}$ and $\hat{V}$:
\begin{equation}\label{}
  \hat{H}_D=\hat{K}+\hat{V}=\left(-i \bm{\alpha} \cdot \bm{\nabla}+m\beta \right)  +\left(-\bm{\alpha} \cdot  e\bm{A}( \bm{r},t) \right). \\
\end{equation}
$\hat{U}(t+\Delta t,t)$ could be approximated by
\begin{equation}\label{}
  \hat{U}(t+\Delta t,t)
  \approx
  \exp \left( -i\frac{\Delta t}{2}\hat{K}\right)
  \exp \left( -i \Delta t \hat{V}(t)\right)
  \exp \left( -i\frac{\Delta t}{2}\hat{K}\right) 
  .
\end{equation}
Notice that $\hat{K}$ and $\hat{V}$ are diagonal in the momentum space and the coordinate space respectively which can be transformed back and forth to each other by the Fourier transformation $\mathcal{F}$ and the inverse Fourier transformation $\mathcal{F}^{-1}$. Therefore, $\phi_{r,s}(t+\Delta t)$ is given by
\begin{widetext}
\begin{equation}\label{}
  \phi_{r,s}(t+\Delta t)=
  \hat{U}(t+\Delta t,t) \phi(t)
  \approx
  \mathcal{F}^{-1}
  \exp \left( -i\frac{\Delta t}{2}\hat{K}\right) \mathcal{F}
  \exp \left( -i \Delta t \hat{V}(t)\right) \mathcal{F}^{-1}
  \exp \left( -i\frac{\Delta t}{2}\hat{K}\right) \mathcal{F} \phi_{r,s}(t).\\
\end{equation}
\end{widetext}
$\phi_{r,s}(t_{out})$ is obtained after $N_t$ steps according to Eq.~(\ref{unitary_decom}).


%

\end{document}